\documentclass[epj,final]{svjour}
\usepackage{latexsym}
\usepackage{graphics}
\begin{document}
\title{Diffusion and Localization of Cold Atoms in 3D Optical Speckle }
\author{A. Yedjour \inst{1,2}\and B.A. van Tiggelen\inst{1}
}                     
%
%
\institute{Laboratoire de Physique et de Mod\'{e}lisation des
Milieux Condens\'{e}s , CNRS /Universit\'{e} Joseph Fourier, BP 166,
F-38042 Grenoble Cedex 9, France
\email{Bart.van-Tiggelen@grenoble.cnrs.fr} \and Laboratoire de
Physique des Plasmas, des Mat\'{e}riaux Conducteurs, et leurs
Applications, Department of Physics, Universit\'{e} des Sciences et
de la Technologie d'Oran, USTO, Oran 31000, Algeria}

\date{Received: date / Revised version: date}
%
\abstract{ In this work we re-formulate and solve the self-consistent
theory for localization to a Bose-Einstein condensate expanding in a
3D optical speckle. The long-range nature of the fluctuations in the
potential energy, treated in the self-consistent Born approximation,
make the scattering strongly velocity dependent, and its
consequences for mobility edge and fraction of localized atoms have
been investigated numerically.
} 
\maketitle

\section{Introduction}

Anderson localization is by now a phenomenon that has been widely
investigated, both theoretically and experimentally, and for many
different kinds of waves, from electrons, to electromagnetic waves,
ultrasound \cite{pht1}, and cold atoms \cite{pht2}. To understand
Anderson localization and to provide quantitative predictions for
experiments, many tool models have been proposed, among which the
Anderson model is undoubtedly the best known. This model describes a
noninteracting and electron, tightly bound to the nucleus, but
capable to tunnel to nearby atoms. Already in the celebrated 1958
paper \cite{pwa}, Anderson demonstrated how this model highlights
the role of dimensionality. A genuine mobility edge only occurs in
dimensions larger than 2 \cite{kramer}. For classical waves, a few
observations on 3D localization have been reported \cite{died,page}.

The tight binding model is highly relevant to understand 3D
dynamical localization of cold atoms \cite{dynamic}. However, is not
appropriate to describe localization of many other waves, where the
starting point is much more a diffuse, extended motion, rather than
a tightly bound state. Localization of electromagnetic waves in 3D
disordered media for instance, is much more a diffusion problem than
a problem of tunneling to nearest neighbors. The scaling theory of
localization \cite{g4}, as well as the Thouless criterion
\cite{thou}, both deal with conductance, and thus use  the diffuse
motion as a starting point.

The self-consistent theory, first formulated by Vollhardt and
W\"{o}lfle in 1981 for 2D electron conductivity \cite{vw}, was the
first work that explicitly calculated how quantum corrections affect  the
classical "Drude" picture, to make way for localization. Despite its evident perturbational
nature, the theory has been
 successful because it provides a microscopic picture of
finite-size scaling, reproduces the Ioffe-Regel criterion for the
mobility edge in 3D, and locates the mobility edge of the
tight-binding model quite accurately \cite{kroha}. The aim of the
present work is to revisit and apply this theory to the localization of cold
atoms in optical speckle.

The first experiments on 1D cold atom localization have been carried
out recently \cite{coldatomloc}. The atoms are released from a BEC
and subsequently expand in a potential energy landscape created by
optical speckle, supposed free of mutual interactions. Both theory
\cite{1Dquasi} and experiment have revealed the presence of a
quasi-mobility edge in 1D. Atoms with velocities $v > \hbar /m \xi $
($\xi$ is the correlation length of the disorder) can hardly be
scattered because this would imply a momentum transfer larger than
$\hbar/\xi$ which the random speckle cannot support. As a result the
localization length is infinite in the Born approximation, though
finite and large when all orders are taken into account. This
somewhat surprising result highlights the impact of long-range correlations in 1D. In higher dimensions, small angle scattering can
still  lead to small enough momentum transfer to be transferred to the speckle, even for large velocities, so
that this quasi-mobility edge does not occur. Yet,
correlations are expected to affect localization, since the potential field
sensed by the atom strongly depends on its velocity, and strong forward scattering
is not favorable for localization to occur. In addition,
near the 3D mobility edge the disorder is necessarily large so that
the spectral function of the atoms is not strongly peaked near
energies $E = {p^2}/{2m}$, neither has it a Lorentzian
broadening.

\section{Self-consistent Born Approximation}

In the following we consider the scattering of a noninteracting atom
with energy $E$ and momentum $\mathbf{p}$ from a disordered
potential $V(\mathbf{r})$. Two properties  are specific for an
optical potential. Firstly, the fluctuations $\delta V(\mathbf{r})$
are determined by the optical intensity and not by the complex
field. This means that they are not Gaussian but rather Poisonnian.
 As a result,  the two-point correlation will in
principle not be sufficient to describe the full scattering
statistics. Secondly, the correlation function, given by
$\langle\delta V(\mathbf{r})\delta V(\mathbf{r}')\rangle = U
\mathrm{sinc}^2 (\Delta r/\xi)$, is long range. Here $U= \langle
V\rangle^2$ depends on the average optical intensity. It is not
difficult to see that the Born approximation breaks down at energies
$E \preceq U/E_\xi$ \cite{kuhn}, with $E_\xi= \hbar^2/2m \xi^2$ an
important energy scale related to correlations. As usual we expect
matter localization to occur at small energies, near the band edge
of the spectrum. Another consequence of the long-range correlations
is that scattering strongly depends on the De Broglie wave length
and thus on the velocity of the atom. This makes it impossible to
define a mean free path $\ell$ in the usual way, that is from the
exponential decay of the ensemble-averaged Green function \cite{pr}.

In the following  we shall cope with the second problem. The
ensemble-averaged Green function is written in terms of a complex
self-energy as $G(E,k)=[E-p^2/2m-\Sigma(E,p) ]^{-1}$ \cite{pr}. We
shall apply the Self-consistent Born Approximation (SCBA) according
to which the complex self-energy $\Sigma(E,p)$ of the atom is
calculated from \cite{scba}
\begin{equation}\label{scba}
    \Sigma(E,k) = \sum_{\mathbf{k}'}  \frac{U(\mathbf{k}-\mathbf{k}')}{E-k'^2 - \Sigma(E,k')}
\end{equation}
In this equation, $\sum_\mathbf{k} \equiv \int
\mathrm{d}^3\mathbf{k}/(2\pi)^3$, and from now,  all energies,
including $\Sigma(E,p)$, are expressed in units of the energy scale
$E_\xi$. Momenta are expressed as $\mathbf{p}=\hbar \mathbf{k}$ with
the De Broglie wave number $\mathbf{k}$ expressed in units of
$1/\xi$. In Eq.~(\ref{scba}), $U(\mathbf{k}-\mathbf{k}')$ represents
the structure function associated with the speckle correlation,
which determines the angular profile in single scattering
\cite{kuhn}. The SCBA is convenient because its imaginary part
expresses the generalized optical theorem in single scattering
\cite{pr}. In addition, it avoids the bound state at negative energies predicted by the first
 Born approximation. In Ref.~\cite{anna} the SCBA was solved analytically for
cold atoms and  zero-range correlations.

Equation~(\ref{scba}) has been solved by iteration, with spline
interpolation between 500 points $0 < k_n < 3$. The angular integral
can be performed analytically. Typically 10-20 iterations have been
necessary to ensure good convergence. In Figure 1 we show real and
imaginary part of $ \Sigma(E,k)$ for $U/E_\xi^2=1$ and $E=0$, and
compare it to the first order Born approximation (FBA) applied in
Ref.~\cite{kuhn}. As a realistic experimental reference we take
$^{87}$Ru-atoms released from a BEC with chemical potential $\mu=
219\, h$ Hz into an optical speckle with correlation length
$\xi=0,26$ $\mu$m. This reveals that $\mu$ and $E_\xi$ are equal
energy scales in typical experiments. Equivalently, the De Broglie wavelength and the
correlation length are competing length scales, $\lambda/2\pi \geq
\xi$. To discriminate "trivial trapping " in deep random potential wells
from ''genuine" Anderson localization, experimentalists wish
to arrange the experiment such that the typical kinetic energies
$\mu$ are somewhat larger than the typical fluctuations $\sqrt{U}$
in the potential energy \cite{coldatomloc}. In that case, $U/E_\xi^2
\geq 1$. We will comment on this choice later.

\begin{figure}[h]
\resizebox{1.0\columnwidth}{!}{%
  \includegraphics{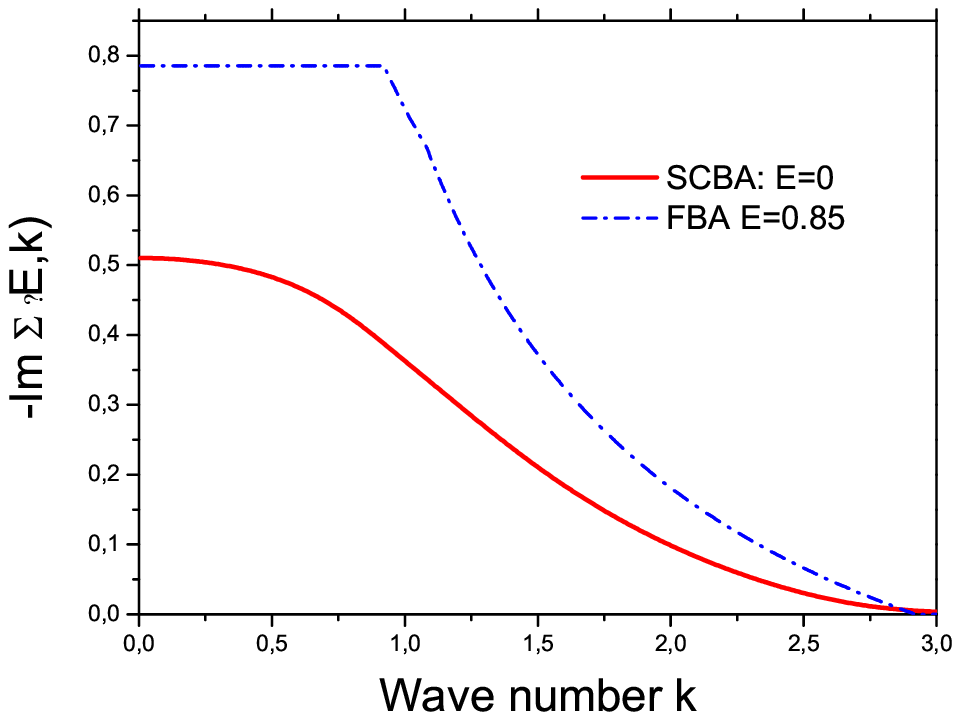}
}
\end{figure}
\begin{figure}[h]
\resizebox{1.0\columnwidth}{!}{%
  \includegraphics{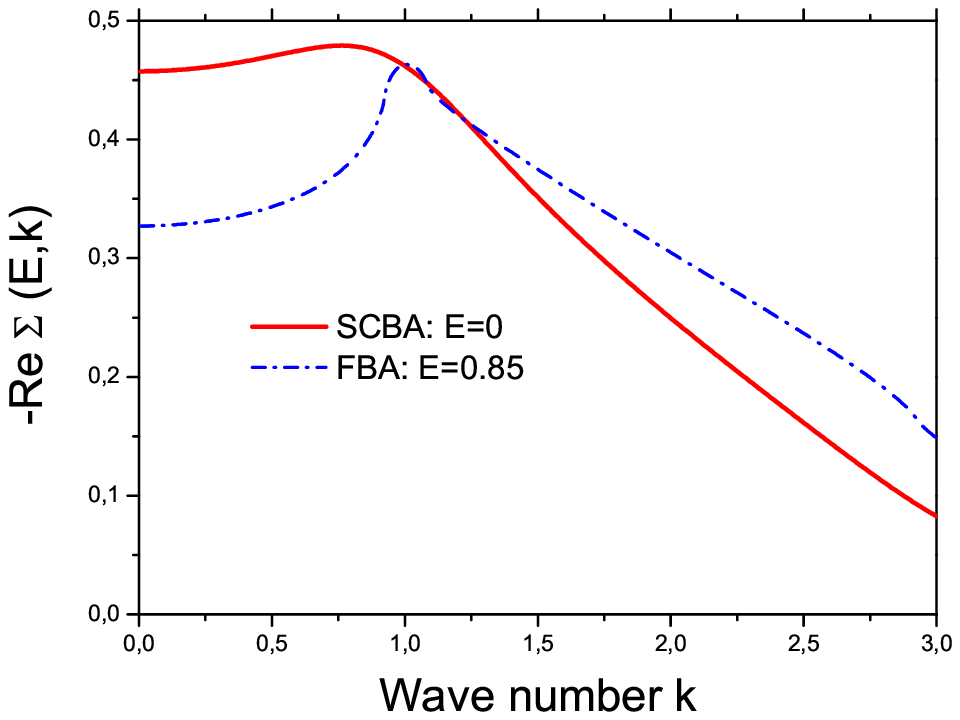}
} \caption{Imaginary (top) and real part (bottom) of the self energy
(in units of $E_\xi$) of an atom in a speckle potential with
$U/E_{\xi}^2=1$, as a function of wave number (in units of $1/\xi$)
for an energy $E=0$, calculated in the self-consistent Born
approximation (SCBA). The dashed line denotes the first Born
approximation (FBA). To compensate for the shift in the band edge
($E_b/E_\xi=-0.85$) predicted by the SCBA, the FBA has been
evaluated at $E/E_\xi=0.85$. }\label{scba1}
\end{figure}

Figure 1 (top) shows $-\mathrm{Im }\, \Sigma(0,k)$ to be nonzero
only for $k< 3$.  It is also seen that the FBA significantly
overestimates the amount of scattering. The real part $\mathrm{Re
}\, \Sigma(0,k)$ is clearly negative. This shifts the band edge of
the energy spectrum to $E_b =-0.85$. The energy spectrum  has a
typical lower bound $ -\sqrt{U}$, but the SCBA always locates the band
edge at somewhat larger energies. The SCBA does probably not treat the (small) density of states near
$E \approx -\sqrt{U}$ very well, where sharply localized Lifshits-type states are likely to occur.
Figure 2 shows the distribution of wave
numbers at energy $E=0$, expressed by the spectral function $S(E,k)
\sim - \mathrm{Im}\, G(E,k)$. It is a rather broad distribution,
with a tail extending to $k=2$. This is important, since we will see
in the next section that atom transport is quite sensitive to large
momentum transfers, involving large $k$-vectors. At low energies,
relevant for localization, $\hbar/\xi$ has become the typical
momentum of an atom, and $E_\xi$ the typical energy. At larger
energies $E=3$, the spectral function behaves normally, i.e.
strongly peaked near $k=\sqrt{E}$. The peak is smaller because we
did not conclude the geometric $4\pi k^2$ surface factor in phase
space, so to highlight its weight at small $k <1 $ (''slow atoms")
for later purposes.

\begin{figure}[h]
\resizebox{1.0\columnwidth}{!}{%
  \includegraphics{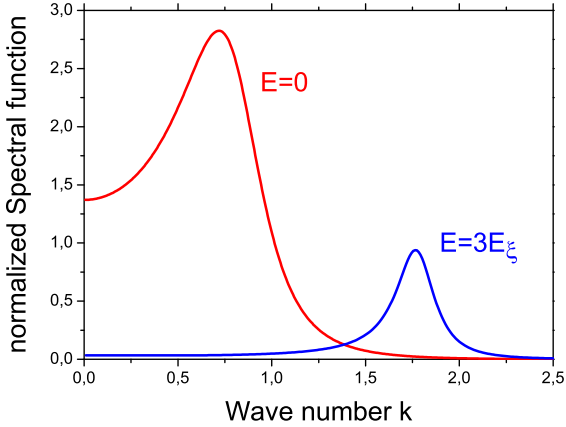}
} \caption{Spectral function, normalized to the total density of
states at energy $E$, calculated in the SCBA, for $E=0$ (red) and
$E=3$ (blue) and
 for $U/E_{\xi}^2=1$. At  small energies it has a large weight at
 $k=0$ (very cold atoms) and extends up to $k=2$.
}\label{spectral}
\end{figure}

\section{Bethe-Salpeter equation}

We proceed with the calculation of the diffusion constant of the
cold atoms, and the possible presence of a mobility edge where it
vanishes. With that information we will find how many atoms will be
localized. The idealized model we consider is schematically drawn in
Figure 3.
\begin{figure}[h]
\resizebox{1.0\columnwidth}{!}{%
  \includegraphics{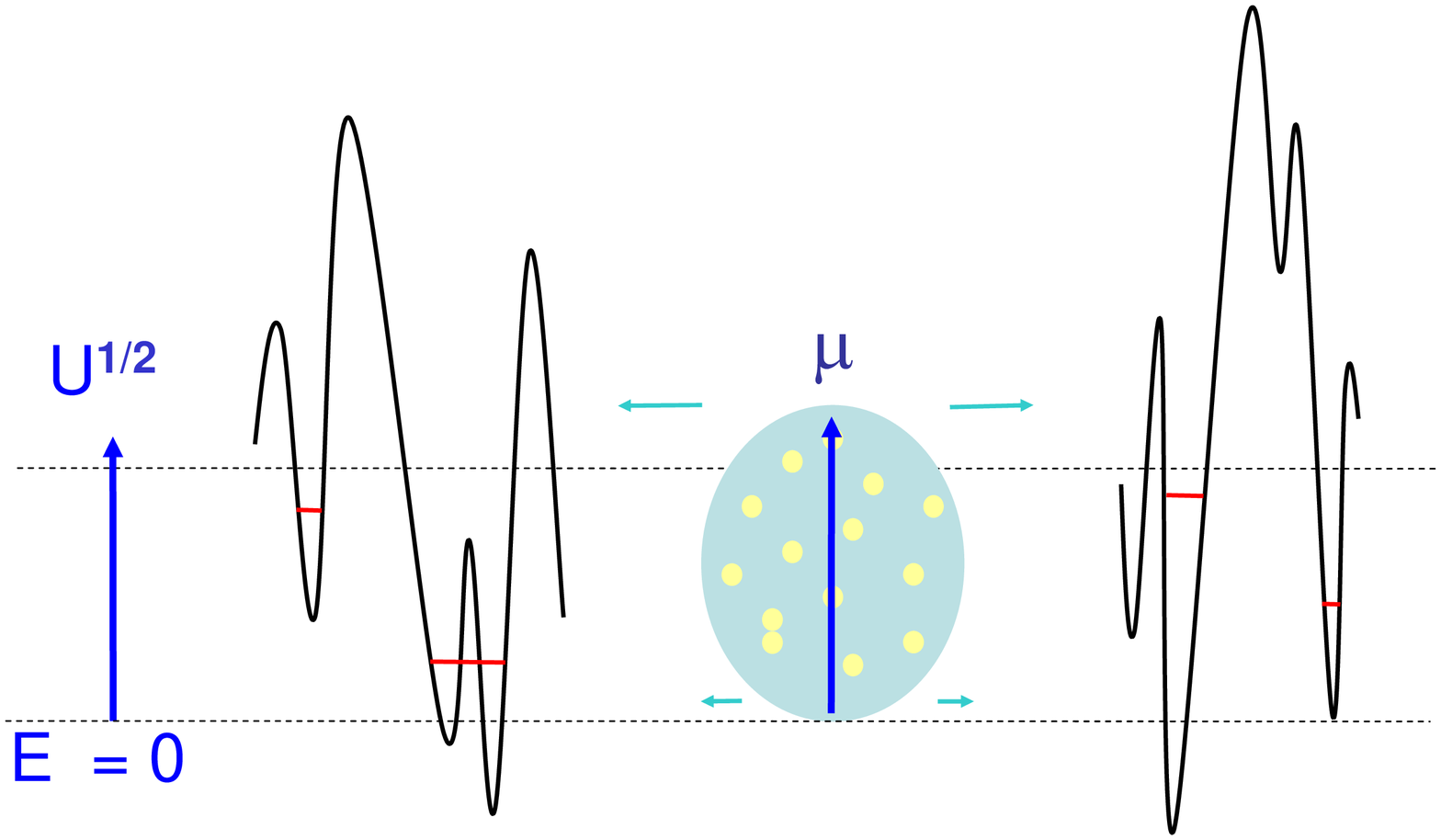}
} \caption{Model of the atomic expansion. The BEC releases
noninteracting atoms with kinetic energies between $0$ and its
chemical potential $\mu$, that penetrate the disordered speckle
potential, with average and its typical fluctuations equal to
$\sqrt{U}$. In this work localized states (red) are found only for energies below the average.}\label{plot}
\end{figure}

The Bethe-Salpeter equation is a rigorous equation for the
two-particle Green function \cite{mahan}. In phase space this object
is written as $\Phi_{\mathbf{kk}'}(E,t, \mathbf{r})$, which is
readily interpreted as the``quantum probability density" for an atom
with velocity $\hbar\mathbf{k}/m$ to travel, during the time $t$,
from position $\mathbf{r}=0$  to position $\mathbf{r}$ and to
achieve the velocity $\hbar\mathbf{k}'/m$. Its Fourier-Laplace
transform with space-time is written as
$\Phi_{\mathbf{kk}'}(E,\Omega, \mathbf{q})$. It has two fundamental
properties, namely reciprocity, $\Phi_{\mathbf{kk}'}(E,\Omega,
\mathbf{q})= \Phi_{\mathbf{k'k}}(E,\Omega, \mathbf{q})$, and
normalization. The last property can be expressed by,

\begin{equation}\label{norm}
    \sum_\mathbf{k} \Phi_{\mathbf{kk}'}(E,\Omega, \mathbf{q}=0) = \frac{-2\mathrm{Im}\,  G(E,k')}{-i\Omega}
\end{equation}
In particular, when also integrating over all energies,

\begin{eqnarray}\label{norm2}
    \int_{-\infty}^\infty \frac{dE}{2\pi} \sum_\mathbf{k,k'} \Phi_{\mathbf{kk}'}(E,\Omega, \mathbf{q}=0)
   &=& \nonumber \\  \frac{1}{-i\Omega}
     \int_{-\infty}^\infty {dE}
    \frac{-1}{\pi}\mathrm{Im}\,  G(E,k) &=& \frac{1}{-i\Omega}
\end{eqnarray}
 The last equality, needed for later purposes, follows from a general sum rule of the spectral
 function \cite{mahan}. This identity guarantees that the total quantum
 probability for the atom to be somewhere, with some velocity and with some energy, is conserved, and equal to one.
The Bethe-Salpeter can be re-written as a quantum-kinetic equation
for $\Phi_{\mathbf{kk}'}(E,\Omega, \mathbf{q})$. The conservation of
quantum probability guarantees the existence of a hydrodynamic
diffusion pole. We shall express this as

\begin{equation}\label{da}
    \Phi_{\mathbf{kk}'}(E,\Omega, \mathbf{q}) =\frac{2}{\sum_\mathbf{k} -\mathrm{Im} \, G(E,k)}
 \, \, \frac{\phi(E,\mathbf{k},\mathbf{q})
\phi(E,\mathbf{k}',\mathbf{q})}{-i\Omega + D(E) q^2}
    \end{equation}
where

$$\phi(E,\mathbf{k},\mathbf{q})= -\mathrm{Im}\, G(E,k) - i(\mathbf{k}\cdot
\mathbf{q})F(E,k) + \mathcal{O}(q^2). $$

This expression states that the distribution of atoms in $k$-space
at given energy $E$ is essentially governed by the spectral
function, with a small correction that supports a current. The front
factor in Eq.~(\ref{da}) is imposed by the normalization condition~(\ref{norm2}).
The $k$-integral of $-\mathrm{Im} \, G(E,k)$ is
recognized as  ($\pi$ times) the density of states (DOS) per unit
volume $\rho(E)$. With the correct normalization, we can set
$\Omega=0$. The still unknown function $F(p)$ follows from
\cite{mahan}

\begin{eqnarray}\label{fp}
    F(E,k) &=& |G(E,k)|^2 -\frac{\partial \mathrm{Re} \, G(E,k)}{\partial
    k^2}  \nonumber \\
    &+& |G(E,k)|^2 \sum_{\mathbf{k'}} \frac{\partial \mathrm{Re} \, G(E,k')}{\partial
    k'^2} \frac{\mathbf{k}\cdot
    \mathbf{k}'}{k^2}U_{\mathbf{kk}'}(E,0) \nonumber \\
    &+&  |G(E,k)|^2 \sum_{\mathbf{k'}} F(E,k') \frac{\mathbf{k}\cdot \mathbf{k}'}{k^2}U_{\mathbf{kk}'}(E,0)
\end{eqnarray}
The irreducible vertex  $U_{\mathbf{\mathbf{kk}}'}(E,\mathbf{q})$
generalizes the function $U(\mathbf{k}-\mathbf{k}')$ defined in the
first section to all interference contributions in multiple
scattering. Once we have solved for $F(E,k)$, the diffusion constant
follows from the Kubo-Greenwood formula \cite{mahan},

\begin{equation}\label{KG}
    D(E)=\frac{\hbar}{m} \frac{2}{3}\frac{1}{\pi \rho(E)} \sum_\mathbf{k} k^2 F(E,k)
\end{equation}
Note that $D(E)$ is determined by the \emph{fourth} moment of the
distribution $F(E,k)$, which puts a large weight on ''fast" atoms.
The order of magnitude of the diffusion constant is governed by the
ratio $\hbar/m$ of Planck's constant and the mass of the atom, the
second factor being dimensionless and of order unity at low
energies. For $^{87}$Ru,  $\hbar/m \approx 1800 $
$\mu$m$^2$/s. The third term in Eq.~(\ref{fp}) can be transformed
using the exact Ward identity,

\begin{eqnarray}\label{ward}
    \Sigma(E, \mathbf{k}+\frac{1}{2}\mathbf{q}) - \Sigma^*(E, \mathbf{k}-\frac{1}{2}\mathbf{q})
    = {\sum_\mathbf{k'}} U_{\mathbf{kk}'}(E,\mathbf{q}) \nonumber \\ \times \left(
    G(E, \mathbf{k'}+\frac{1}{2}\mathbf{q}) - G^*(E, \mathbf{k'}-\frac{1}{2}\mathbf{q})
    \right)
\end{eqnarray}
If this identity is developed linearly in $\mathbf{q}$, and inserted
into Eq.~(\ref{fp}) , we obtain,

\begin{eqnarray}\label{fp2}
    F(E,k) &=&  F_0(E,k)  + \delta_\mathbf{q} U(E,k) \nonumber \\
    &+& |G(E,k)|^2 \sum_{\mathbf{k'}} F(E,k') \frac{\mathbf{k}\cdot
\mathbf{k}'}{k^2}U_{\mathbf{kk}'}(E,0)
\end{eqnarray}
with

$$F_0(E,k)\equiv |G(E,k)|^2 \left( 1 + \frac{\partial \mathrm{Re} \, \Sigma(E,k)}{\partial
    k^2}\right) -\frac{\partial \mathrm{Re} \,
G(E,k)}{\partial
    k^2} $$

$$\delta_\mathbf{q} U\equiv  |G(E,k)|^2\sum_{\mathbf{k}'} \mathrm{Im}\,
    G(E,k') \left( 2i\frac{\mathbf{k}}{k^2}\cdot\frac{\partial}{\partial\mathbf{q}}\right) U_{\mathbf{kk}'}(E,\mathbf{q})
    $$.

\bigskip

Three levels of analysis exist. First, in the Drude
approximation, one neglects all contributions form the BS-equation
and one adopts $F(E,p) = F_0(E,p)$, including the wave number
derivatives. The Drude diffusion constant $D_d(E)$ can be used to
define a dimensionless Ioffe-Regel type parameter from the relation
$D_d(E)= \frac{1}{3} (2\hbar /m) k \ell$. Hence,

\begin{equation}\label{ir}
    k\ell \equiv \frac{\sum_\mathbf{k}k^2 F_0(E,k)}{\sum_\mathbf{k} -\mathrm{Im}\, G(E,k)}
\end{equation}
For a short-range correlation, the self-energy is independent of $k$
so that $F_0(E,k)= 2 \mathrm{Im}^2 \, G(E,k)$, and this definition of $k\ell$
coincides with the usual one in terms of $\-\mathrm{Im} \,
\Sigma(E)$ \cite{pr}. For low energies we found in Figure 3 that typically  $k
\approx 1/\xi$ . If we anticipate that $k\ell \approx
1$ near the mobility edge, we conclude that the mean free path is roughly equal to the correlation length.

\begin{figure}[h]
\resizebox{1.0\columnwidth}{!}{%
 \includegraphics{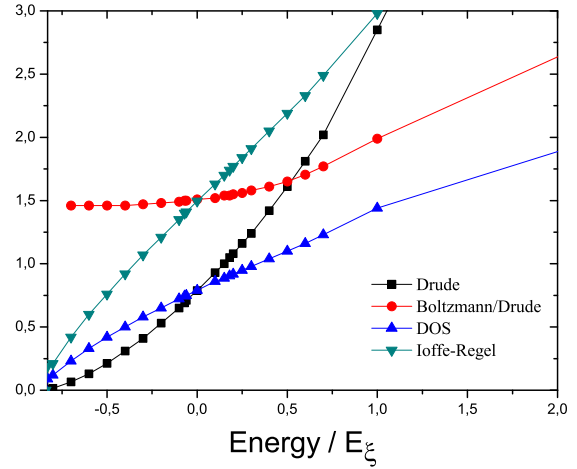}
 }
 \caption{Solution of the BS equations~(\ref{KG}) and (\ref{fp2}) in the Boltzmann
approximation, for a disorder amplitude $U/E_{\xi}^2 =1$. Shown
as a function of energy are the Drude conductivity $\pi
\rho(E)D_d(E)$, the Ioffe-Regel parameter~(\ref{ir}),
the DOS that vanishes at the band edge $E/E_\xi=-0.83$,  and the
ratio of Boltzmann and Drude conductivity. Note that $2/3$ times the
Ioffe-Regel parameter equals the Drude diffusion constant, expressed
in units of $\hbar/m$. }
\end{figure}
\begin{figure}[h]
\resizebox{1.0\columnwidth}{!}{%
  \includegraphics{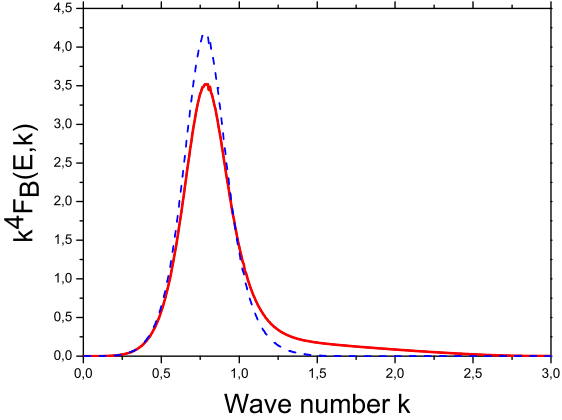}
}  \caption{The function $k^4F_B(E,k)$, solution of the BS-equation
for the energy $E/E_\xi=-0.06$ in the Boltzmann approximation. The blue dashed line compares it to
the Drude approximation $k^4F_0(E,p)$ defined in Eq.~(\ref{fp2}),
rescaled properly such that both $k$-integrals equal
$\sigma_B/\sigma_0$ }\label{boltz}
\end{figure}

The Drude approximation is popular in electron - impurity scattering
but clearly inadequate when the scattering is strongly anisotropic,
as for the optical speckle. In the Boltzmann approximation we adopt
$U_{\mathbf{kk}'}=U(\mathbf{k-k}')$, i.e. the structure function
associated with the optical disorder. Being a function of
$\mathbf{k}-\mathbf{k}'$ only, it follows that $\delta_\mathbf{q}
U =0$. Different results are summarized  in Figure 4. At low
energies Boltzmann and Drude diffusion constant typically differ by
a factor $1.5$ as was obtained by Kuhn etal \cite{kuhn}
on the basis of the FBA. For atom energies $E > E_\xi$ the Boltzmann
diffusion constant rapidly rises since only strong forward scattering can occur. In the region $E=0$, the
Ioffe-Regel parameter takes values of the order of $1.5$. We
infer from Figure 5 that the solution $F_B(E,p)$ is roughly a
rescaling of the Drude Ansatz, for which  the current is dominated by
atoms with velocities $v=0.83 \hbar/m \xi$. Nevertheless, a
non-negligible fraction of atoms faster than $v=1.5 \hbar/m \xi$
contributes to the current (note that $\hbar/m \xi \approx 7 $ mm/s
for the set-up with Rubidium described above).
\bigskip

We finally consider constructive interferences, and
add the most-crossed diagrams to the BS-equation in the spirit of
the self-consistent theory of localization. Any observation $D <
D_B$ must be attributed to constructive interferences. It is
well-known that, by reciprocity, these diagrams can be constructed
from the solution $\phi(E,\mathbf{k},\mathbf{q})$ of the BS-equation
by removing the incoming and outgoing Green's functions (indicated
by a hat), and by time-reversing the bottom line \cite{eplbart}:
$U^{MC}_{\mathbf{kk}'}(E,\mathbf{q})=
\widehat{\Phi}_{\frac{1}{2}(\mathbf{k}-\mathbf{k}'+\mathbf{q})\,
\frac{1}{2}(\mathbf{k}'-\mathbf{k}+\mathbf{q}) }(E,
\mathbf{k}+\mathbf{k}')$. Using Eq.~(\ref{da}) the two procedures
lead to

\begin{eqnarray}\label{MC}
 &\,&  \widehat{ \Phi}_{\mathbf{kk}'}(E)=    \frac{2}{\sigma
   (E)\mathbf{q}^2} \times \nonumber \\
   &\ & \left[ \mathrm{Im} \Sigma(E,k) \mathrm{Im}
   \Sigma(E,k') \right.
+ \left. i\mathbf{q}\cdot(I_{kk'}(E)\mathbf{k} +
   I_{k'k}(E)
   \mathbf{k}')\right]
   \nonumber \\
   &\, & U^{\mathrm{MC}}_{\mathbf{kk}'}(E,\mathbf{q}) =
   \frac{2}{\sigma(E)(\mathbf{k}+\mathbf{k}')^2}
   \times \nonumber \\
&\  & \left[   \mathrm{Im}^2 \Sigma(E,\frac{1}{2}\Delta k ) +
I_{\Delta k /2, \Delta k /2}(E)
   i(\mathbf{k}+\mathbf{k}')\cdot\mathbf{q} \right]
\end{eqnarray}
We abbreviated  $I_{kk'} = F(E,k)\mathrm{Im}\,
\Sigma(E,k')/|G(E,k)|^2$, $\Delta k$  $= |\mathbf{k} -\mathbf{k}'| $
and introduced  $\sigma(E)=\pi\rho(E)D(E)$, the equivalent of
DC-conductivity in electron conduction. We now face the more
complicated task of solving Eqs.~(\ref{fp2}) and (\ref{KG})
simultaneously with $U_{\mathbf{kk}'}=U(\mathbf{k}-\mathbf{k}')+
U^{\mathrm{MC}}_{\mathbf{kk}'}$, and of finding out if its
extrapolation to small energies leads to a mobility edge where
$\sigma(E)=0$. This constitutes a "self-consistent" problem for the
entire function $F(E,k)$, rather than just for its fourth moment,
the DC-conductivity.

We first observe that for most-crossed diagrams $\delta_\mathbf{q} U
\neq 0$. This term does not appear in standard self-consistent
theory \cite{vw}, which relies on moment expansion. Note that it
also features "self-consistently" the function $F(k)/\sigma$,
just like the second term in Eq.~(\ref{fp2}). As can be induced from
Eq.~(\ref{MC})  the singularity at $\mathbf{k}=-\mathbf{k}'$ that
generates the (weak) localization is partly compensated by the
factor $\mathbf{k}+\mathbf{k}'$. We shall therefore ignore it here
as well.

The self-consistent equations~(\ref{fp2}),(\ref{KG}) and (\ref{MC})
can be solved almost analytically when the self-energy is assumed
$k$-independent, typically true for zero-range correlations. Without
more details we mention that the mobility edge then occurs at
$k\ell=1.122$ \cite{eplloc}. Quite convenient is that, even when
scattering extends to infinite wave numbers, our theory does not
require an ad-hoc cut-off to eliminate short wave paths that diverge
in approximate theories \cite{kuhn,economou}. For the speckle
correlation, Figure 6 gives the result of the exact numerical
solution, obtained by iteration and spline interpolation. This
method worked satisfactorily until close to ($\Delta E \approx
E_\xi/5$) the mobility edge, where the most-crossed diagrams give a
diverging contribution. Before that happens, the strong forward
scattering of a single scattering competes heavily with the
reduction induced by weak localization.

\begin{figure}[h]
\resizebox{1.0\columnwidth}{!}{%
 \includegraphics{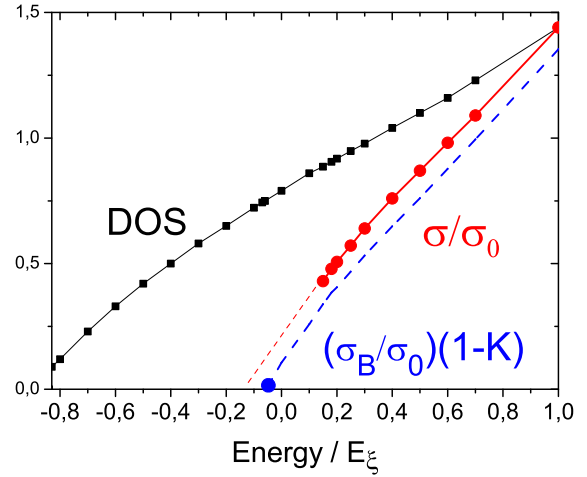} }
 \caption{Solution of the BS equation~(\ref{fp2}) with inclusion of
the most-crossed diagrams for a disorder amplitude $U/E_{\xi}^2 =1$.
Shown in red is the ratio of conductivity and Drude conductivity for energies
$E> 0.15 E_\xi$ for which our iteration converged. The blue dashed
line relies on an approximation discussed in the text and locates
the mobility at $E=-0.06 E_\xi$.}
\end{figure}
\begin{figure}[h]
\resizebox{1.0\columnwidth}{!}{%
  \includegraphics{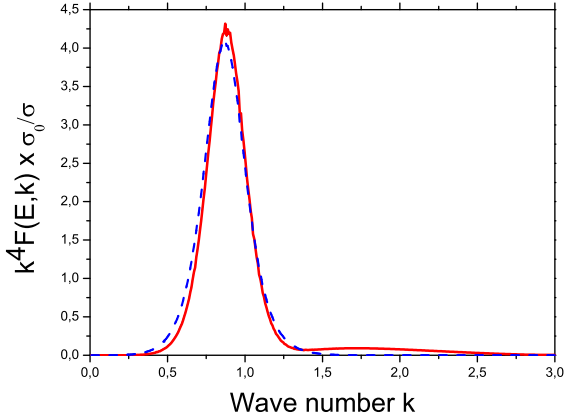}
 } \caption{The function $k^4F(E,k)$
(in red)), solution of the BS-equation for the energy
$E/E_\xi=0.15$, compared to the Drude approximation $k^4F_0(E,p)$
defined in Eq.~(\ref{fp2}) (blue dashed line). The first has been
rescaled by the factor $\sigma_0/\sigma$ such that the $k$-integrals
of both are equal to one. }\label{VW}
\end{figure}

To find the location of the mobility edge we shall use the following
approximation. In Figure 7 we see that at $E/E_\xi=0.15$  the
solution $F(E,k)/\sigma $ is closely approximated by
$F_0(E,k)/\sigma_0$ (both have their fourth moment normalized to
one). This equivalence is physically reasonable since it implies
that all atoms with energy $E$ undergo the same reduction in
diffusion, but keep the same velocity distribution as found in the Drude picture.
If we insert
$F_0(E,k)$ in the left hand side of Eq.~(\ref{fp2}) and integrate
over $k$ we can derive the simple relation $\sigma/\sigma_B \approx
1-K(E)$ that is reminiscent of standard self-consistent theory
\cite{vw}. The parameter $K$ is found to be,

\begin{eqnarray}\label{K}
    K(E) = -\frac{4}{3 \sigma_d^2(E)}\sum_{\mathbf{kk}'}
    F_0(E,k')\frac{\mathbf{k}\cdot
    \mathbf{k}'}{(\mathbf{k}+\mathbf{k}')^2} \times \nonumber \\
    \mathrm{Im}^2\Sigma\left(E,\frac{1}{2}|\mathbf{k}-\mathbf{k}'|
    \right)
    |G(E,k)|^2
\end{eqnarray}

\begin{figure}[h]
\resizebox{1.0\columnwidth}{!}{%
 \includegraphics{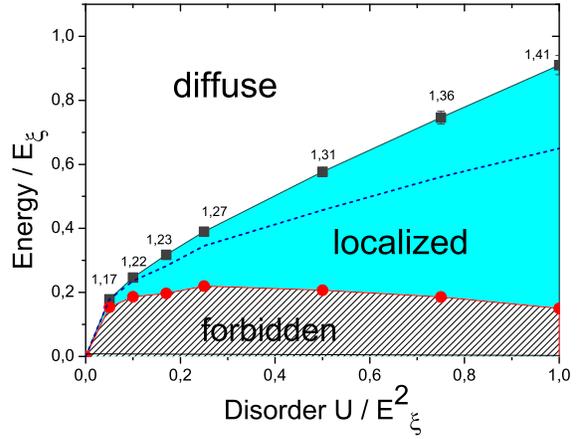} }
\caption{Band edge and mobility edge as a function of the disorder
strength $U$, all expressed in the energy $E_\xi$. Contrary to the
former figures the energy scale has been shifted by the average
potential $\sqrt{U}$ of the potential energy so that a direct
comparison can be made to the chemical potential $\mu > 0$, and the
criterion $E_c -E_b \approx 0.5 U/E_\xi$ found by Ref.\cite{kuhn}
(blue dashed). The numbers around the mobility edge reveal the small
variation of the Ioffe-Regel parameter~(\ref{ir}). The exclusion of small energies is likely to be an artifact
of the  SCBA and we expect strongly localized Lifshits-type states with small DOS.}\label{energies}
\end{figure}
For $U/E^2_\xi=1$, this approximation locates the mobility edge
($K=1$) at $E/E_\xi=-0.06$ ($0.94$ when we shift energies over
$\sqrt{U}$ as in Fig 8). Upon inspecting the numerically exact
solution in Figure 8, we suspect that the real mobility edge is
located somewhat lower, near $E/E_\xi=-0.1$.  For smaller disorder
we found that the approximation becomes better. It is in principle
possible to obtain numerically the function $F(E,p)/\sigma$ as
$\sigma \rightarrow 0$ and calculate more precisely the location of
mobility edge, but this is beyond the scope of this paper. In Figure
8 we show the different energies for different strengths of the
disorder. If we apply the criterion found in Ref~\cite{kuhn},
$k\ell_B =0.95$ (dashed line in Figure 8), localization would occur
at smaller energies, around $E/E_\xi=-0.35$ ($0.65$ in Fig. 8) for
$U/E^2_\xi=1$. This approach  expresses the general trend well but
is clearly somewhat pessimistic, probably because  their choice of
the ad-hoc wave number cut-off to calculate the most-crossed diagrams
 underestimates $K$. Note that all localized states occur \emph{below} the average of the potential
 landscape. Unlike in 1D, we find
 no regime with
 atoms fast enough to traverse the potential barriers, but to become localized purely
 by constructive interferences, without the assistance of tunneling.

A final important question is how many atoms will be localized,
given an initial velocity distribution that is determined by the
expansion of the BEC after eliminating the trap. We emphasize
that according to the scenario sketched in Fig. 3, the chemical potential
$\mu$ of the BEC does not represent the energy of the atom inside the disordered potential, but rather the distribution of incident kinetic energies. If this
distribution is denoted by $\phi_\mu(\mathbf{k})$, it follows from
Eq.~(\ref{da}) that the fraction  of localized atoms, regardless of
their final velocity or position, is given by

\begin{eqnarray}\label{nloc}
    \int_{E_b}^{E_c} \frac{dE}{2\pi} \sum_\mathbf{k,k'} \Phi_{\mathbf{kk}'}(E,t,
    \mathbf{q}=0)\phi_\mu(\mathbf{k}') \nonumber \\ =
     \int_{E_b}^{E_c}{dE} \sum_\mathbf{k'}
    \frac{-1}{\pi}\mathrm{Im}\,  G(E,k')\phi_\mu(\mathbf{k}')
\end{eqnarray}

Castin and Dum \cite{yvan} showed that after the free expansion $\phi_\mu(k) \sim 1-k^2/k^2_\mu$
and zero for $k
> k_\mu$, where the maximum wave number $k_\mu = \sqrt{\mu/E_\xi}$.
The fraction of localized atoms is thus determined by the number of
microscopic states below the mobility edge whose kinetic energies
are smaller than $\mu$. In this discussion it is convenient to make
the zero of the energy scale the same for $\mu$ and $E$, as was
already done already  in Figure 8, shifting the localized region to
positive energies. In present (1D) experiments is $U/E_\xi^2=1$ and
$\mu \approx E_\xi$ which is only slightly above the 3D mobility
edge. One might perhaps expect most atoms to be localized. Our
calculations clearly show that the distribution of atom energies is
quite different from $\phi_\mu(\mathbf{k})$. This
(normalized) distribution $F(E)$ is given by the wave number
integral in Eq.~(\ref{nloc}) and is shown in Figure 9. For $\mu \ll
E_\xi$ it reduces to the spectral function at $k=0$, independent of
$\phi_\mu(k)$. Even for small $\mu$, many atoms achieve energies $E
> E_c$ ($40\, \%$ for $U/E^2_\xi=1$) and are delocalized. This
number agrees  well with predictions based on zero-range
correlations in which case $45 \, \%$ was found to be localized for
$\mu \ll E_c$ \cite{anna}. The fraction of delocalized atoms further
decreases as the chemical potential rises (Figure 10), with only
$35\, \%$ localized for $\mu = E_\xi$. Even if we choose $ \mu
> \sqrt{U}$, the atoms that localize have energies $E < \sqrt{U}$.

\begin{figure}[h]
\resizebox{1.0\columnwidth}{!}{%
 \includegraphics{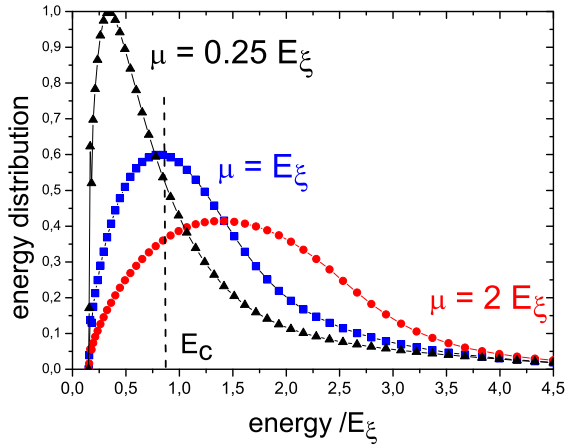} }
 \caption{Energy distributions of the atoms inside the
disordered speckle potential, for different chemical potentials of
the BEC from which they were released and $U/E_\xi^2=1$. They all
exhibit a tail of relatively fast atoms ($E > E_\xi$) that extends
beyond the mobility edge $E_c$. }
\end{figure}
\begin{figure}[h]
\resizebox{1.0\columnwidth}{!}{%
 \includegraphics{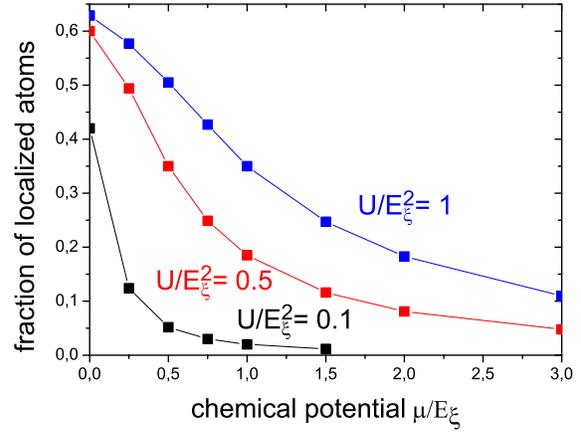} }
 \caption{Fraction of localized atoms
as a function of chemical potential, for different amplitudes of the
disorder.}\label{nloc}
\end{figure}

In conclusion, we have calculated the phase diagram for localization
of cold atoms in a 3D speckle potential, using the self-consistent
Born approximation and the self-consistent theory of localization.
The mobility edge is characterized by a Ioffe-Regel type parameter
that varies between $1.2$ and $1.4$. Depending on the chemical
potential of the BEC, typically $35\, \%$ to $60\, \%$ of the atoms
are localized. The self-consistent Born approximation deals already
much better with the long-range correlations and the broadness of
the spectral function than the first Born approximation, but does
not discriminate between different statistics of the disorder. Yet,
the mobility edge of the tight-binding model is known to depend on
that \cite{kroha}. It would be very interesting to apply a recently
proposed method \cite{schindler} to calculate the self-energy of the
atoms more precisely. The  theory presented in this work can then be used
straightforwardly to find the mobility edge.

\begin{acknowledgement}
We would like to thank Sergey Skipetrov  for help and discussions.
\end{acknowledgement}

\end{document}